\documentclass[aps,pra,twocolumn,showpacs,superscriptaddress]{revtex4-1}
\usepackage{bm}
\usepackage{mathrsfs}
\usepackage{amsmath}
\usepackage{amssymb}
\usepackage{graphicx}
\usepackage{amsfonts}
\usepackage{amsthm}
\usepackage{color}
\usepackage{dcolumn}
\usepackage{txfonts}
\usepackage{epsfig}

\begin{document}
\title{Relations between Heat Exchange and R\'{e}nyi Divergences}
\author{Bo-Bo Wei}
\email{Corresponding author: bbwei@szu.edu.cn}
\affiliation{School of Physics and Energy, Shenzhen University, Shenzhen 518060, China}

\begin{abstract}
In this work, we establish an exact relation which connects the heat exchange between two systems initialized in their thermodynamic equilibrium states at different temperatures and the R\'{e}nyi divergences between the initial thermodynamic equilibrium state and the final non-equilibrium state of the total system. The relation tells us that the various moments of the heat statistics are determined by the Renyi divergences between the initial equilibrium state and the final non-equilibrium state of the global system. In particular the average heat exchange is quantified by the relative entropy between the initial equilibrium state and the final non-equilibrium state of the global system. The relation is applicable to both finite classical systems and finite quantum systems.
\end{abstract}
\maketitle

\section{Introduction}
Work and heat are two central themes of equilibrium thermodynamics \cite{Adkins1983}. However, for finite systems, the work and the heat exchange are random numbers and both of them have a distribution. In an intriguing development Jarzynski \cite{Jarzynski1997} found that for a finite classical system prepared in the thermodynamic equilibrium state the work in a non-equilibrium process is related to the equilibrium free energy differences between the thermodynamic equilibrium states for the final and initial control parameters. The Jarzynski equality establishes deep connections between the equilibrium free energy differences and the work in a non-equilibrium process\cite{PNAS2001,Science2005,Nature2005,EPL2005,PT2005,PRL2006,PRL2007}. Afterwards, the Jarzynski equality has been demonstrated to hold in finite quantum mechanical systems based on the two projective measurements \cite{arXiv2000a,arXiv2000b,PRL2003,Talker2007,Kim2015}. Since the discovery of the Jarzynski equality, the investigation of various fluctuation relations in non-equilibrium thermodynamics has become an active research field \cite{RMP2009,Jar2011,RMP2011}. Recently the author and his collaborator found that \cite{Wei2017a,Wei2017b,Wei2017c} the dissipated work which is defined as the work minus the free energy difference is related to the R\'{e}nyi divergences between two microscopic states in the forward and reversed dynamics. This relation links the dissipation in non-equilibrium thermodynamics to the R\'{e}nyi divergences in information theory and has recently been verified experimentally in a superconducting qubit system \cite{Fan2017}.

In 2004, Jarzynski and W\'{o}jcik discovered that the heat exchange between two systems $A$ and $B$ which are initialized in the thermodynamic equilibrium states at different temperatures $T_A$ and $T_B$ respectively satisfies the following fluctuation relation \cite{Jarzynski2004}
\begin{eqnarray}
\langle e^{-\Delta\beta Q}\rangle=1.
\end{eqnarray}
Here $\Delta\beta=\beta_B-\beta_A$ with $\beta=1/T$ being the inverse temperature. For more developments on the exchange fluctuation relations, one may refer to \cite{RMP2011}. The central result of this paper is the following exact relation about the heat exchange between two systems $A$ and $B$,
\begin{eqnarray}\label{central}
\langle \Big(e^{-\Delta\beta Q}\Big)^z\rangle =e^{(z-1)S_{z}[\rho(0)||\rho(\tau)]},
\end{eqnarray}
where $z$ is an arbitrary real number, $Q$ is the heat exchange between $A$ and $B$, the angular bracket on the left side denotes an ensemble average over all realizations of the heat exchange process and $S_{z}[\rho(0)||\rho(\tau)]\equiv\frac{1}{z-1}\ln[\text{Tr}[\rho(0)^{z}\rho(\tau)^{1-z}]]$ is the order-$z$ R\'{e}nyi divergence \cite{Renyi1961,Erven2014,Beigi2013,Lennert2013} between the initial equilibrium state of the total system $\rho(0)$ and the final non-equilibrium state of the total system at time $\tau$, $\rho(\tau)$. Equation \eqref{central} is applicable for both finite classical mechanical systems and finite quantum mechanical systems. For classical system, $\rho(0)$ and $\rho(\tau)$ appeared in Equation \eqref{central} should be understood as the corresponding phase space density and the trace is replaced by an integral over the entire phase space of the total system. While for quantum mechanical system, $\rho(0)$ and $\rho(\tau)$ in Equation \eqref{central} should be understood as the quantum density matrices of the total system. Equation \eqref{central} connects a macroscopic quantity, the heat exchange between two systems,  and a microscopic quantity, the R\'{e}nyi divergences of two microscopic states.

Because the definition of heat exchange in classical systems and in quantum systems are different \cite{RMP2011}, we shall discuss derivation of Equation \eqref{central} for classical systems and quantum systems respectively.

\section{Heat Exchange for Classical Systems}
First of all, we discuss the heat exchange between two classical systems which are initialized respectively in their thermodynamic equilibrium state at different temperatures.

\subsection{Distribution of Heat Exchange in Classical Systems}
Consider two classical systems $A$ and $B$ and their Hamiltonians are given by $H_A(X^A)$ and $H_B(X^B)$ respectively. Here $X^A$ denotes a phase space point in the system $A$ and $X^B$ denotes a phase space point in the system $B$ and $X=(X^A,X^B)$ labels a point in the phase space of the global system. Now let us describe the heat exchange process in the classical system. \\
1.~At $t=0$, we initialize two classical systems in the thermodynamic equilibrium states at temperatures $T_A$ and $T_B$ respectively. For simplicity, we assume $T_A>T_B$. Then the initial state of the whole system is given by
\begin{eqnarray}
\rho(X_0;0)&=&\frac{e^{-(\beta_AH_A(X_0^A)+\beta_BH_B(X_0^B))}}{Z_AZ_B}.
\end{eqnarray}
Here $X_0=(X_0^A;X_0^B)$ is a phase space point in the global system, $\beta_A=1/T_A$, $\beta_B=1/T_B$, $Z_A=\int dX_0^A e^{-\beta_AH_A(X_0^A)}$ is the partition function of the system $A$ and the partition function of system $B$ is $Z_B=\int dX_0^B e^{-\beta_BH_B(X_0^B)}$.\\
2.~We measure energies of the two systems $A$ and $B$ respectively and the outcomes are $H_A(X_0^A)$ and $H_B(X_0^B)$ with the corresponding probability $\rho(X_0;0)$.\\
3.~Let the two systems in thermal contact with each other for a time duration $\tau$ which is arbitrary and the total Hamiltonian including the interactions is
\begin{eqnarray}
\mathcal{H}=H_A(X^A)+H_B(X^B)+H_{AB}(X^A;X^B).
\end{eqnarray}
4.~We separate the systems $A$ and $B$ and then measure energies of the two systems respectively and the results are  $H_A(X_1^A)$ and $H_B(X_1^B)$ respectively. Here $X_1=(X_1^A,X_1^B)$ is the phase space point of the global system at time $\tau$ under the dynamics governed by the total Hamiltonian $\mathcal{H}$ when the initial phase space point is $X_0=(X_0^A,X_0^B)$. We assume the interaction $H_{AB}$ is weak such that the energy of the global system is approximately conserved and thus the heat exchange is given by \cite{Jarzynski2004}
\begin{eqnarray}
Q=H_B(X_1^B)-H_B(X_0^B)=H_A(X_0^A)-H_B(X_1^B);
\end{eqnarray}
Thus the probability distribution for heat exchange is given by
\begin{eqnarray}
P(Q)&=&\int dX_0\rho(X_0;0)\delta\left(Q-H_B(X_1^B)+H_B(X_0^B)\right).
\end{eqnarray}

\subsection{Heat Exchange and R\'{e}nyi Divergences in Classical Systems}
Now we are ready to evaluate the generating function of heat exchange,
\begin{eqnarray}
&&\langle\left(e^{-\Delta\beta Q}\right)^z\rangle\nonumber\\
&=&\int dX_0\rho(X_0;0)e^{-z\Delta\beta(H_B(X_1^B)-H_B(X_0^B))},\\
&=&\frac{1}{Z_AZ_B}\int dX_0 e^{-(\beta_AH_A(X_0^A)+\beta_BH_B(X_0^B))}e^{-z\beta_B(H_B(X_1^B)-H_B(X_0^B))}\nonumber\\
&& \times e^{-z\beta_A(H_B(X_0^A)-H_B(X_1^A))},\\
&=&\frac{1}{Z_AZ_B}\int dX_0 \left(e^{-(\beta_AH_A(X_0^A)+\beta_BH_B(X_0^B))}\right)^{1-z}\nonumber\\
&& \times \left(e^{-(\beta_AH_A(X_1^A)+\beta_BH_B(X_1^B))}\right)^{z},\\
&=&\int dX_0 \rho(X_0;0)^{1-z}\rho(X_1;0)^{z},\label{c1}\\
&=&\int dX_1 \rho(X_1;\tau)^{1-z}\rho(X_1;0)^{z},\label{c2}\\
&=&\exp\left[(z-1)S_z\left(\rho(X;0)||\rho(X;\tau)\right)\right].\label{c3}
\end{eqnarray}
Here $\Delta\beta=\beta_B-\beta_A>0$ and $z$ is an arbitrary real number. From Equation \eqref{c1} to \eqref{c2}, we have made use of the Liouville theorem in Hamilton dynamics, which states that the phase space density along a trajectory of the classical system is invariant, which means that $\rho(X_0,0)=\rho(X_1,\tau)$ and also the phase space volume is invariant under Hamilton dynamics $dX_0=dX_1$. From Equation \eqref{c2} to \eqref{c3}, we made use of the definition of R\'{e}nyi divergences or R\'{e}nyi relative entropy \cite{Renyi1961,Erven2014,Beigi2013,Lennert2013}, $S_z(\rho_1||\rho_2)=\frac{1}{z-1}\ln\int dX\rho_1(X)\rho_2(X)$, which is the order-z R\'{e}nyi divergence between two probability distributions $\rho_1$ and $\rho_2$. We thus derived the relation between heat exchange and the Renyi divergences in classical systems,
\begin{eqnarray}\label{centralC}
\langle\left(e^{-\Delta\beta Q}\right)^z\rangle=\exp\left[(z-1)S_z\left(\rho(X;0)||\rho(X;\tau)\right)\right].
\end{eqnarray}
Now we make several remarks on the above equality:\\
1. In the above equality \eqref{centralC}, $z$ is a free parameter and it can take any real values. In the case of $z=1$, we recover
\begin{eqnarray}
\langle e^{-\Delta\beta Q}\rangle=1.
\end{eqnarray}
This is the fluctuation relation for heat exchange first derived by Jarzynski in 2004 \cite{Jarzynski2004}.\\
2. The generating function of heat exchange shall give us the various moments of heat statistics. In particular, the average heat exchange is given by
\begin{eqnarray}
(\beta_B-\beta_A)\langle Q\rangle&=&D\left(\rho(X,\tau)||\rho(X,0)\right),
\end{eqnarray}
where the right hand side is the relative entropy \cite{relativeentropy} between the final non-equilibrium phase space density of the global system $\rho(X,\tau)$ and the initial equilibrium phase space density of the global system $\rho(X,0)$. Furthermore, the higher order moments of the heat exchange is given by
\begin{eqnarray}
\langle Q^n\rangle&=&(\beta_B-\beta_A)^{-n}\int dX\rho(X,\tau)\left(\ln\frac{\rho(X,\tau)}{\rho(X,0)}\right)^n.
\end{eqnarray}
Here $n=1,2,3,\cdots$.\\
3. The Renyi divergences is a valid measure of the distinguishability \cite{Renyi1961,Erven2014,Beigi2013,Lennert2013} and thus the Renyi divergence appears in the equation \eqref{centralC} means that heat exchange is a consequence of non-equilibrium dynamics.\\
4. The equality \eqref{centralC} is valid for any interaction time $\tau$ which is a consequence of Liouville theorem in Hamilton dynamics.

\section{Heat Exchange for Quantum Systems}
In this section, we consider the heat exchange between two quantum systems $A$ and $B$.

\subsection{Distribution of Heat Exchange in Quantum Systems}
The heat exchange between two systems $A$ and $B$ is defined by the following steps:\\
1. The two systems $A$ and $B$ are separately prepared at their own thermodynamics equilibrium states at different temperatures $T_A$ and $T_B$. For simplicity, we assume that $T_A>T_B$. Then the initial state of the whole system is
\begin{eqnarray}
\rho(0)&=&\frac{e^{-\beta_A H_A}}{Z_A}\otimes \frac{e^{-\beta_B H_B}}{Z_B},
\end{eqnarray}
where $\beta=1/T$, $Z_A=\text{Tr}[e^{-\beta_AH_A}]$ is the equilibrium partition function of system $A$ and the partition function of the system $B$ is defined by $Z_B=\text{Tr}[e^{-\beta_BH_B}]$.\\
2. We perform the first projective measurement of the energy $H_A$ and $H_B$ respectively and we assume the outcomes are $E_{n,A}$ and $E_{n,B}$ respectively with the corresponding probability
\begin{eqnarray}
p_n(0)=\frac{e^{-(\beta_A E_{n,A}+\beta_B E_{n,B})}}{Z_AZ_B}.
\end{eqnarray}
Here $H_A|n,A\rangle=E_{n,A}|n,A\rangle$ and $H_B|n,B\rangle=E_{n,B}|n,B\rangle$. At the same time the state of the two system is projected into the corresponding eigenstate $|n\rangle=|n_A,n_B\rangle$. \\
3. We allow the systems $A$ and $B$ to interact for some time $\tau$ and the total Hamiltonian including the interactions is
\begin{eqnarray}
\mathcal{H}=H_A+H_B+H_{AB}.
\end{eqnarray}
After interactions, the state of the global system is $ \mathcal{U}_{0,\tau}|n\rangle $.
Here the time development operator is given by
\begin{eqnarray}
\mathcal{U}_{0,\tau}=\exp\left(-it\mathcal{H}\right)=\exp\left(-it(H_A+H_B+H_{AB})\right).
\end{eqnarray}
4. After interactions for time interval $\tau$, we then separate the systems $A$ and $B$.\\
5. Finally we perform the second projective measurement of the energy $H_A$ and $H_B$ respectively and we assume the outcomes are $E_{m,A}$ and $E_{m,B}$ respectively and the conditional probability for obtaining  $E_{m,A}$ and $E_{m,B}$ is
\begin{eqnarray}
p_{n\rightarrow m}=\left|\langle m|\mathcal{U}_{0,\tau}|n\rangle\right|^2.
\end{eqnarray}
Here $|m\rangle=|m_A,m_B\rangle$ and they satisfy the Schr\"{o}dinger equation, $H_A|m_A\rangle=E_{m,A}|m,A\rangle$ and $H_B|m_B\rangle=E_{m,B}|m,B\rangle$.
We assume the interactions is so weak that approximately the energy of the total system is conserved. Thus the heat exchange between the two systems $A$ and $B$ is
\begin{eqnarray}
Q=E_{n,A}-E_{m,A}\approx E_{m,B}-E_{n,B}.
\end{eqnarray}
Thus the quantum heat exchange distribution is \cite{Jarzynski2004}
\begin{eqnarray}
P(Q)&=&\sum_{m,n}p_n(0)p_{n\rightarrow m}\delta\left(Q-E_{m,B}+E_{n,B}\right),\\
&=&\sum_{m,n}p_n(0)\left|\langle m|\mathcal{U}_{0,\tau}|n\rangle\right|^2\delta\left(Q-E_{m,B}+E_{n,B}\right),\\
&=&\sum_{m,n}\frac{e^{-(\beta_AE_{n,A}+\beta_BE_{n,B})}}{Z_AZ_B}\left|\langle m|\mathcal{U}_{0,\tau}|n\rangle\right|^2\delta\left(Q-E_{m,B}+E_{n,B}\right).\nonumber\\
\end{eqnarray}
The characteristic function of quantum heat exchange is given by the Fourier transform of the distribution of heat exchange,
\begin{eqnarray}\label{cf}
G(u)&=&\int dQP(Q)e^{iuQ},\nonumber\\
&=&\text{Tr}\left[e^{-(\beta_AH_A+\beta_BH_B)}e^{-iuH_B}\mathcal{U}_{0,\tau}^{\dagger}e^{iuH_B}\mathcal{U}_{0,\tau}\right].
\end{eqnarray}
It should be noted that the characteristic function for heat exchange in Equation \eqref{cf} has an analogous expression to the quantum decoherence of a probe spin coupled to a bath \cite{Yang2017} which has been demonstrated to connect to the partition function of the bath in the complex plane of physical parameters \cite{Wei2012,Wei2014,Wei2015,Peng2015,LYExp2015,Wei2017a1,Wei2017b1,Wei2017b1}.

\subsection{Heat Exchange and R\'{e}nyi Divergences in Quantum Systems}
The generating function of heat exchange between $A$ and $B$ are
\begin{eqnarray}
&&\langle \Big(e^{-\Delta\beta Q}\Big)^z\rangle\nonumber\\
&=&\int dQP(Q)e^{-z\Delta\beta Q},\\
&=&\sum_{m,n}\frac{e^{-(\beta_AE_{n,A}+\beta_BE_{n,B})}}{Z_AZ_B}\left|\langle m|\mathcal{U}_{0,\tau}|n\rangle\right|^2e^{-z(\beta_B-\beta_A)(E_{m,B}-E_{n,B})},\\
&=&\sum_{m,n}\frac{e^{-(\beta_AE_{n,A}+\beta_BE_{n,B})}}{Z_AZ_B}\langle m|\mathcal{U}_{0,\tau}|n\rangle \langle n|\mathcal{U}_{0,\tau}^{\dagger}|m\rangle e^{-z(\beta_B-\beta_A)(E_{m,B}-E_{n,B})},\\
&=&\frac{1}{Z_AZ_B}\sum_{m,n}\langle m|\mathcal{U}_{0,\tau}e^{-(1-z)(\beta_AH_A+\beta_BH_B)}|n\rangle \langle n|\mathcal{U}_{0,\tau}^{\dagger}e^{-z(\beta_AH_A+\beta_BH_B)}|m\rangle,\nonumber\\ \\
&=&\frac{1}{Z_AZ_B}\text{Tr}\left[\mathcal{U}_{0,\tau}e^{-(1-z)(\beta_AH_A+\beta_BH_B)}\mathcal{U}_{0,\tau}^{\dagger}e^{-z(\beta_AH_A+\beta_BH_B)} \right],\\
&=&\frac{1}{Z_AZ_B}\text{Tr}\left[\left(\mathcal{U}_{0,\tau}e^{-(\beta_AH_A+\beta_BH_B)}\mathcal{U}_{0,\tau}^{\dagger}\right)^{1-z}\left(e^{-(\beta_AH_A+\beta_BH_B)}\right)^z \right],\\
&=&\text{Tr}\left[\left(\mathcal{U}_{0,\tau}\rho(0)\mathcal{U}_{0,\tau}^{\dagger}\right)^{1-z}\rho(0)^z \right],\\
&=&\text{Tr}\left[\rho(\tau)^{1-z}\rho(0)^z \right],\\
&=&\exp\left[(z-1)S_z\left(\rho(0)||\rho(\tau)\right)\right].
\end{eqnarray}
Here $\rho(\tau)=\mathcal{U}_{0,\tau}\rho(0)\mathcal{U}_{0,\tau}^{\dagger}$. Thus we have derived the relation between heat exchange and the Renyi divergences in quantum mechanical systems,
\begin{eqnarray}\label{centralQ}
\langle \left(e^{-\Delta\beta Q}\right)^z\rangle&=&\exp\left[(z-1)S_z\left(\rho(0)||\rho(\tau)\right)\right].
\end{eqnarray}
We make several remarks on the above relation:\\
1. In the equation \eqref{centralQ}, $z$ is an arbitrary real number and if $z=1$ we get
\begin{eqnarray}
\langle e^{-\Delta\beta Q}\rangle&=&1.
\end{eqnarray}
This is the exchange fluctuation theorem in quantum mechanical systems first derived by Jarzynksi in 2004 \cite{Jarzynski2004}.\\
2. The average heat exchange is given by
\begin{eqnarray}
(\beta_B-\beta_A)\langle Q\rangle&=&D\left(\rho(\tau)||\rho(0)\right).
\end{eqnarray}
Here the right hand side is the relative entropy \cite{relativeentropy} between the final non-equilibrium state $\rho(\tau)$ and the initial equilibrium state of the global system $\rho(0)$. Moveover, the higher order moments of the heat exchange is given by
\begin{eqnarray}
\langle Q^n\rangle&=&(\beta_B-\beta_A)^{-n}\text{Tr}\left[\rho(\tau)\mathcal{T}_n\left(\ln[\rho(\tau)]-\ln[\rho(0)]\right)^n\right].
\end{eqnarray}
Here $n=1,2,3,\cdots$ and $\mathcal{T}_n$ is an ordering operator which sorts that in each term of the binomial expansion $\left(\ln[\rho(\tau)]-\ln[\rho(0)]\right)^n$, $\ln[\rho(\tau)]$ always lies on the left of $\ln[\rho(0)]$.\\
3. The Renyi divergences is a valid measure of the distinguishability \cite{Renyi1961,Erven2014,Beigi2013,Lennert2013} and thus the Renyi divergence appears in the equation \eqref{centralQ} for quantum system means that the heat exchange comes from the non-equilibrium quantum dynamics of the systems due to interactions. \\
4. The equality \eqref{centralQ} is valid for any interaction time $\tau$ which is a consequence of unitarity in quantum dynamics.\\
5. It was proposed that the quantum heat exchange could be measured from the Ramsey interference of a single spin \cite{Goold2014}. Thus our relation \eqref{centralQ} means that we can also measure the family of quantum R\'{e}nyi divergences between the equilibrium state and an out of equilibrium state of a quantum system by Ramsey interference experiment.

\section{Summary}
In summary, we have derived an exact equality which relates the heat exchange between two systems initialized in the thermodynamic equilibrium states at different temperatures and the R\'{e}nyi divergences between the initial thermodynamic equilibrium state and the final non-equilibrium state of the global system. Because the Renyi divergence is a valid measure of distinguishability, the relation implies that the heat exchange comes from non-equilibrium dynamics of two systems. The relation implies that the various moments of the heat statistics are determined by the R\'{e}nyi divergences. In particular, the average heat exchange between two systems initially prepared at different temperatures is quantified by the relative entropy of the initial equilibrium state and the final non-equilibrium state of the global system. Our results are consequences of two assumptions, namely the initial states of the two systems are described by the Gibbs ensemble and the interaction energy in the process of contact is negligible compared to the energy of both systems. The relation is applicable to both finite classical systems and finite quantum systems. Finally it is conceivable that similar relations could be derived for particle exchange between two systems initialized in the thermodynamic equilibrium states with different chemical potentials.

\begin{acknowledgements}
This work was supported by the National Natural Science Foundation of China (Grant Number 11604220) and the Startup Fund of Shenzhen University under (Grant number 2016018).
\end{acknowledgements}

\end{document}